\documentclass[aps, pre, amsfonts, amssymb, amsmath, superscriptaddress, showpacs, reprint]{revtex4-1}
\usepackage{times}
\usepackage{bm}
\usepackage{hyperref}
\usepackage{graphicx}
\usepackage{epsfig} 

\newcommand{\xx}{\mathbf x}
\newcommand{\rr}{\mathbf r}
\newcommand{\vv}{\mathbf v}
\newcommand{\uu}{\mathbf u}
\newcommand{\bb}{\mathbf b}
\newcommand{\ee}{\mathbf e}
\newcommand{\jj}{\mathbf j}
\newcommand{\zz}{\mathbf z}
\newcommand{\ff}{\mathbf f}

\newcommand{\pp}{\mathbf p}
\newcommand{\BB}{\mathbf B}

\newcommand{\D}{\mathrm{d}}
\newcommand{\inj}{_{\scriptsize \mbox{inj}}}
\newcommand{\tot}{^{\scriptsize \mbox{tot}}}

\begin{document}

\title{Acceleration of Particles in Imbalanced Magnetohydrodynamic Turbulence}
\author{Bogdan Teaca}
\email{bogdan.teaca@coventry.ac.uk}
\affiliation{Applied Mathematics Research Centre, Coventry University, Coventry CV1 5FB, United Kingdom}
\affiliation{Max-Planck/Princeton Center for Plasma Physics and Max-Planck-Institut f\"ur Plasmaphysik, D-85748 Garching, Germany}
\affiliation{Max-Planck f\"ur Sonnensystemforschung, Max-Planck-Str. 2, D-37191 Katlenburg-Lindau, Germany}
\author{Martin S. Weidl}
\affiliation{Max-Planck/Princeton Center for Plasma Physics and Max-Planck-Institut f\"ur Plasmaphysik, D-85748 Garching, Germany}
\author{Frank Jenko}  
\affiliation{Max-Planck/Princeton Center for Plasma Physics and Max-Planck-Institut f\"ur Plasmaphysik, D-85748 Garching, Germany}
\author{Reinhard Schlickeiser}  
\affiliation{Institut f\"ur Theoretische Physik, Lehrstuhl IV: Weltraum- und Astrophysik, Ruhr-Universit\"at Bochum, D-44780 Bochum, Germany}
\begin{abstract}
The present work investigates the acceleration of test particles, relevant to the solar-wind problem, in balanced and imbalanced magnetohydrodynamic turbulence (terms referring here to turbulent states possessing zero and nonzero cross helicity, respectively).  These turbulent states, obtained numerically by prescribing the injection rates for the ideal invariants, are evolved dynamically with the particles. While the energy spectrum for balanced and imbalanced states is known, the impact made on particle heating is a matter of debate, with different considerations giving different results. By performing direct numerical simulations, resonant and non-resonant particle accelerations are automatically considered and the correct turbulent phases are taken into account. For imbalanced turbulence, it is found that the acceleration rate of charged particles is reduced and the heating rate diminished. This behavior is independent of the particle gyroradius, although particles that have a stronger adiabatic motion (smaller gyroradius) tend to experience a larger heating.  
\end{abstract}
%Magnetohydrodynamics in plasma dynamics and flow; Astrophysical plasma; Acceleration of particles in interplanetary space;  
%
%
\maketitle

%%%%%%%%%%%%%%%%%%%%%%%%%%%%%%%%%%%%%%%%%%%%%%%%%%%%%%%%%%%%
%%%%%%%%%%%%%%%%%%%%%%%%%%%%%%%%%%%%%%%%%%%%%%%%%%%%%%%%%%%%
%  Introduction
%%%%%%%%%%%%%%%%%%%%%%%%%%%%%%%%%%%%%%%%%%%%%%%%%%%%%%%%%%%%
{\em Introduction. ---} The slow and fast streams in the solar wind represent good examples of balanced and imbalanced states (differing by the level of cross-helicity; to be defined below) of magnetohydrodynamic (MHD) turbulence, respectively. This formalism captures the large-scale fluctuations, compared to the proton thermal gyroradius. Although a kinetic approach is needed for the treatment of scales smaller than the proton gyroradius \cite{Bale:2005p1273, Alexandrova:2009p1396}, where the interaction of kinetic Alfv\'{e}n waves \cite{Howes:2011p1370} and electron heating of the solar wind \cite{Howes:2008p1524} become important, the self-organization of turbulent structures remains predominantly a large-scale effect, determined by fluid-like dynamics.

In MHD turbulence, the conservation of cross-helicity for the ideal systems represents a dynamical constraint of interest, as it is the quantity that leads to a balanced or imbalanced state of MHD turbulence. While the scaling of the energy spectra for these states has received a lot of attention in recent years \cite{Boldyrev:2006p196, Perez:2009p48, Wicks:2011p1270}, less effort was given to understand the impact on particle acceleration and heating due to the different arrangement of structures. Although it is commonly accepted that the energy transfer rate is reduced in imbalanced MHD turbulence (for which the cross helicity is nonzero) compared to balanced turbulence \cite{Dobrowolny:1980p1525, Hossain:1995p1552, Chandran:2009p760}, the issue of whether particle heating is similarly dependent on the degree of imbalance has been answered differently by different authors. Quasi-linear theory, in which the diffusion of particle position and momentum in MHD turbulence is quantified via Fokker-Planck coefficients \cite{2002cra..book.....S,Hauff:2010p1239}, predicts a strong dependence \cite{Dung:1990p1557,Dung:1990p1533}, although it was recently suggested that the perpendicular heating rate of ions in the solar wind may not be significantly affected by imbalance \cite{Chandran:2010p1532}.

In this context, two general questions arise: how do dynamical constraints on a macroscopic level, responsible for the self-organization process of turbulent structures, affect the acceleration of charged particles, and how does this behavior depend on the gyroradius? In the current paper we will numerically investigate the problem using non-relativistic test particles accelerated by the fully self-consistent electromagnetic field in time-dependent MHD fields. This study provides insight into the problem and, to a certain degree, links the fluid and kinetic approaches.

%%%%%%%%%%%%%%%%%%%%%%%%%%%%%%%%%%%%%%%%%%%%%%%%%%%%%%%%%%%%
%%%%%%%%%%%%%%%%%%%%%%%%%%%%%%%%%%%%%%%%%%%%%%%%%%%%%%%%%%%%
%  The MHD equations
%%%%%%%%%%%%%%%%%%%%%%%%%%%%%%%%%%%%%%%%%%%%%%%%%%%%%%%%%%%%
{\em Basic equations. ---} 
The incompressible MHD equations need to be formulated in terms of two dynamical quantities. These quantities can be either the plasma velocity ($\uu$) and the self-consistent magnetic field ($\mathcal{\mathbf{b}}$, hereafter being expressed in Alfv\'{e}n velocity units $\bb \rightarrow \mathcal{\mathbf{b}} / \sqrt{\rho \mu_0}$, where $\rho$ is the constant mass density) or the Elsasser variables \cite{Elsasser:1950p424}, defined as $\zz^\pm\!=\!\uu \pm \bb$. The Elsasser representation can be seen as the nonlinear scattering of counter-propagating Alfv\'{e}n waves of fluctuations $\zz^\pm$, traveling along the large-scale structures of the magnetic filed (not necessarily just along a mean magnetic field $\BB_0$). In terms of Elsasser variables, the incompressible MHD equations can be written as
\begin{align}
\frac{\partial \zz^\pm}{\partial t} &\!=\! - (\zz^\mp \mp \BB_0) \cdot \! \nabla \zz^\pm \!+\! \nu^+ \nabla^2 \zz^\pm \!+\! \nu^- \nabla^2 \zz^\mp \!+\! \ff^\pm  \!-\! \nabla p \;, \label{elss}
\end{align}
where $\nu^\pm=(\nu\pm\eta)/2$, with $\nu$ being the kinematic fluid viscosity and $\eta$ being the magnetic diffusivity. The total (hydrodynamic and magnetic) pressure field ($p$) is an auxiliary variable that enforces the incompressibility condition for the velocity field. Since the magnetic field is intrinsically zero-divergent, the zero divergence conditions of the Elsasser fields read $\nabla \cdot \zz^\pm=0$. Finally, the divergence-free fields $\ff^\pm$ correspond to a known external forcing mechanism used to reach a stationary state of the system.

%%%%%%%%%%%%%%%%%%%%%%%%%%%%%%%%%%%%%%%%%%%%%%%%%%%%%%%%%%%%
%%%%%%%%%%%%%%%%%%%%%%%%%%%%%%%%%%%%%%%%%%%%%%%%%%%%%%%%%%%%
%  Balanced and imbalanced turbulence
%%%%%%%%%%%%%%%%%%%%%%%%%%%%%%%%%%%%%%%%%%%%%%%%%%%%%%%%%%%%
{\em Balanced and imbalanced turbulence. ---} 
For ideal MHD fluctuations, three quadratic invariants exist: total energy ($\mathcal  E=\frac{1}{2}\langle  \uu(\xx)\cdot \uu(\xx) +  \bb(\xx)\cdot \bb(\xx) \rangle_\xx$), cross-helicity ($\mathcal  H^c=\langle  \uu(\xx)\cdot \bb(\xx) \rangle_\xx$), and magnetic helicity ($\mathcal  H^m= \langle  \bf{a}(\xx)\cdot \bb(\xx) \rangle_\xx$, where $\bb=\nabla\times \bf{a}$). Here, $\langle \cdots \rangle_\xx$ refers to spatial averages. Ideal MHD refers to the case of absent external forces and dissipative effects ($\nu =\eta =0$) in Eq.~(\ref{elss}). For dissipative systems, ideal invariants lead to constant spectral fluxes that redistribute invariant quantities between different scales and link the scaling of the dynamical fields, a fact evident from the Politano-Pouquet equations \cite{Politano:1998p783}. This link is responsible for the self-organized nature of MHD turbulence. In the Elsasser representation, the cross-helicity and total energy information is contained in the definition of two ideal invariants $\mathcal  E^+=\langle  E^+(\xx) \rangle_\xx=\frac{1}{4}\langle  \zz^+(\xx)\cdot \zz^+(\xx) \rangle_\xx$ and $\mathcal  E^-=\langle  E^-(\xx) \rangle_\xx=\frac{1}{4}\langle  \zz^-(\xx)\cdot \zz^-(\xx) \rangle_\xx$, referred to as pseudo-energies. While their sum will obviously give the total energy in the system ($\mathcal  E^++ \mathcal  E^-=\mathcal  E$), their difference ($\mathcal  E^+- \mathcal  E^-=\mathcal  H^c$) measures the preference of the system to generate one type of wave over the other. This led to the names {\em balanced turbulence} (for $\mathcal  H^c=0$) and {\em imbalanced turbulence} $(\mathcal  H^c\neq 0$) being used in the literature. The cross-helicity level, defined as
\begin{align}
&\sigma^c = \frac{\mathcal H^c}{\mathcal E} \equiv \frac{\mathcal E^+-\mathcal E^-}{\mathcal E^++ \mathcal E^-}\;, \label{cross}
\end{align}
represents a better way to quantify MHD states exactly, as $\sigma^c\in[-1,1]$. A value close to $\pm 1$ denotes strongly imbalanced turbulent states, while in a state of $\sigma^c=\pm1$ no nonlinear interaction can take place (Alfv\'{e}n states).
At a point-wise level, $\sigma^c\ne0$ denotes a preference in the generation of the same sign for $\cos [\angle(\uu, \bb)]$ \cite{Milano:2001p1551, Matthaeus:2008p53}, which affects the structure of the electric field \cite{Breech:2003p1546} that particles experience.

%%%%%%%%%%%%%%%%%%%%%%%%%%%%%%%%%%%%%%%%%%%%%%%%%%%%%%%%%%%%
%%%%%%%%%%%%%%%%%%%%%%%%%%%%%%%%%%%%%%%%%%%%%%%%%%%%%%%%%%%%
%  The MHD stationary states
%%%%%%%%%%%%%%%%%%%%%%%%%%%%%%%%%%%%%%%%%%%%%%%%%%%%%%%%%%%%
{\em MHD stationary states. ---} 
Numerically, we employ the {\sc turbo} code \cite{Teaca:2009p628} to solve the MHD equations. Since the mean field value is on the order of the rms magnetic fluctuations ($B_0\sim \delta b$, with $\delta b = \langle \bb^2(\xx)\rangle_\xx^{1/2}$) we use a cubic $(2\pi)^3$ domain, with periodic boundary conditions, discretized using $512^3$ grid points. For comparison, results obtained in the absence of a mean magnetic field are indicated explicitly. The pseudo-spectral method used is consistent with a direct-numerical-simulation (DNS) approach, such that $k_{\max}\ell_K\sim1.25$, where $\ell_K$ is the smallest turbulent scale of the system estimated as the Kolmogorov scale $\ell_K = ( {\nu^3}/{\varepsilon})^{1/4}$ with $\varepsilon$ being the energy dissipation level in the system ($\varepsilon \equiv\mathcal D^{\tot} =\langle \nu \nabla^2 \uu^2 +\eta \nabla^2 \bb^2 \rangle_\xx$) and $\nu=\eta=6.6\times 10^{-4}$. For the time integration, a third-order Runge-Kutta method is used with an adaptive time step determined by a Courant-Friedrichs-Lewy condition. The numerical aliasing effects are suppressed by a two-thirds dealiasing method \cite{Patterson:1971p965}.

%%%%%%%%%%
\begin{figure}[t]
\begin{center}
\includegraphics[width = 0.48\textwidth]{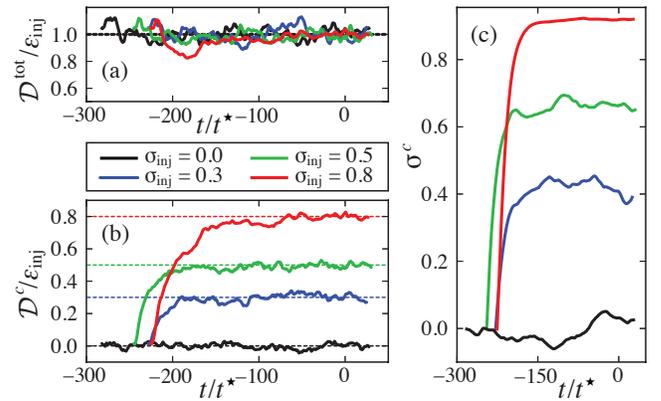}
\end{center}
\vskip -5mm
\caption{(color online). a) the evolution of the total (sum of kinetic and magnetic) energy dissipation $\mathcal D^{\tot}$; b) the evolution of the cross-helicity dissipation $\mathcal D^{c}=\langle (\nu+\eta) \nabla^2 \uu\cdot\bb \rangle_\xx$, normalized to the energy injection rate $\varepsilon_{\inj}$ and c) the time evolution of the cross-helicity level. Test particles are injected at $t/t^\star=0$.}  
\label{fig_evolution}
\end{figure}
%%%%%%%%%%

%%%%%%%%%%
\begin{figure*}[tb]
\begin{center}
\includegraphics[width = 0.98\textwidth]{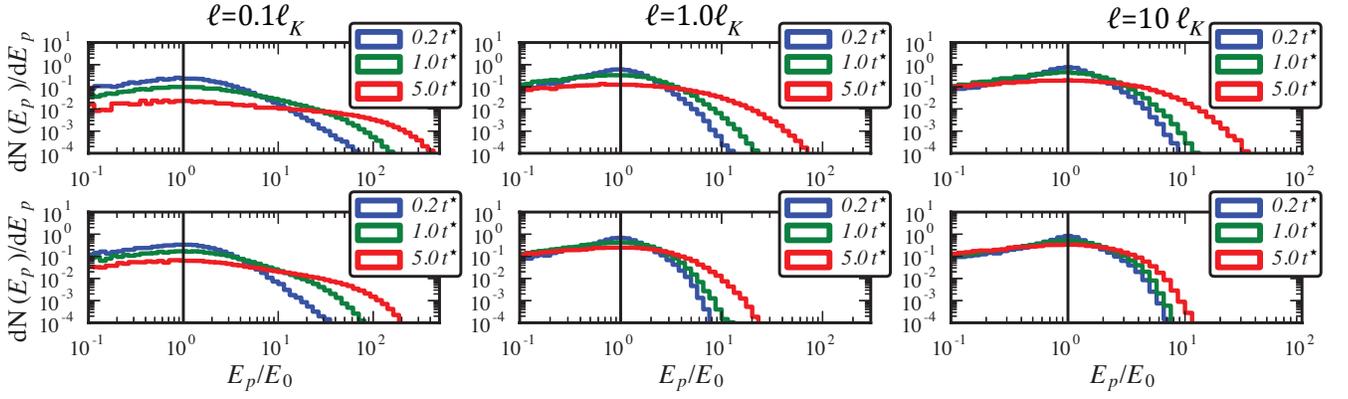}
\end{center}
\vskip -5mm
\caption{(color online). Histograms of the kinetic energy distributions of an initially mono-energetic particle ensemble with $E_p(0) = v_A^2 / 2$, for balanced turbulence ($\sigma=0$) on top row and strongly imbalanced turbulence ($\sigma=0.9$) on bottom row. }  
\label{fig_histograms}
\end{figure*}
%%%%%%%%%%%

To achieve balanced or imbalanced stationary states, we use a forcing mechanism that is local in Fourier space and acts in the same manner on all the modes within a wavenumber shell $s_{f}=[2.5, 3.5]$. Selecting a large number of forced modes ensures that no anisotropy effect is induced by the forcing mechanism. The forces are defined on a helical-mode basis \cite{Waleffe:1993p974} and use the injection rates of the total energy (here $\varepsilon_{\inj}=0.1$), cross-helicity level ($\sigma_{\inj}$) and magnetic helicity (taken as zero for all cases) as control parameters. The forcing method used here imposes the dissipation levels for the energy (Fig.~\ref{fig_evolution}-a) and cross-helicity (Fig.~\ref{fig_evolution}-b) in the stationary regime, without modifying the phases of the fields, see \cite{Teaca:2011p1338} for details. This ensures that no change is made in the type of turbulent structures present. We should note that $\ff^+$ and $\ff^-$ force the two types of Alfv\'{e}n waves. Thus, an imbalanced case can only be obtained for $\ff^+\ne\ff^-$, for which $\ff^b=\frac{1}{2}[\ff^+-\ff^-]\ne0$. In this study, along the control case of balanced turbulence ($\sigma_{\inj}=0$), we consider the imbalanced cases given by $\sigma_{\inj}=\{0.3, 0.5, 0.8\}$ and for which $\sigma^c=\{0.4, 0.6, 0.9\}$, respectively (Fig.~\ref{fig_evolution}-c). The time is normalized by $t^\star$, here taken as the Alfv\'en time.

%%%%%%%%%%%%%%%%%%%%%%%%%%%%%%%%%%%%%%%%%%%%%%%%%%%%%%%%%%%%
%%%%%%%%%%%%%%%%%%%%%%%%%%%%%%%%%%%%%%%%%%%%%%%%%%%%%%%%%%%%
%  Particle tracking
%%%%%%%%%%%%%%%%%%%%%%%%%%%%%%%%%%%%%%%%%%%%%%%%%%%%%%%%%%%%
{\em Particle tracking. ---}
In order to investigate how the stochastic acceleration of charged particles differs for various degrees of cross helicity, we evolve the trajectories of test-particles in parallel with the MHD simulations. The test-particles are injected after the MHD turbulence has attained a steady-state and are evolved using a Newtonian representation,
\begin{align}
\label{eqnMotion}
\frac{{\mbox d}\rr}{{\mbox d}t}  = \vv,\ \ \ \
\frac{{\mbox d}\vv}{{\mbox d}t} = \frac{1}{\ell} \Big{[} \ee(\rr) + \vv \times [ \BB_0 + \bb(\rr) ]\Big{]},
\end{align}
where $\rr(t)$ and $\vv(t)$ are the position and the velocity of a particle at time $t$. The electric field is computed from Ohm's law for resistive MHD, $\ee = \eta \jj - \uu \times (\BB_0 + \bb)$, with $\jj=\nabla \times \bb$. The coupling parameter $\ell=(\frac{q}{m}\sqrt{\rho\mu _0})^{-1}$ represents the particle's charge to mass ratio ($q/m$), takes into account the use of Alfv{\'e}nic units by the electromagnetic fields and has units of length. Intuitively, it can be seen as the Larmor radius of a particle that moves in a constant magnetic field with the perpendicular velocity equal to the local Alfv\'en velocity ($v_A$) of the magnetic field. Numerically, a cubic-spline interpolation on the fields and an implicit fourth-order Runge-Kutta solver with adaptive step-size control are employed to advance the particle trajectories (see Ref. \cite{Lalescu:2013p1433} for details).

Throughout this work, we consider five particle species with 50,000 test-particles per species. All particles start at a random initial position, with a random direction for the velocity and with an initial velocity equal to the Alfv\'en velocity, $v_0=v_A$ (here $v_A^2=B_0^2+\delta b^2$). The five particle species have $\ell=\{0.1, 0.3, 1, 3, 10\}\times \ell_K$. Taking the values in units of the Kolmogorov scale length is a choice that allows us to compare the initial maximal gyroradii to the smallest turbulent scale.

%%%%%%%%%%%%%%%%%%%%%%%%%%%%%%%%%%%%%%%%%%%%%%%%%%%%%%%%%%%%
%%%%%%%%%%%%%%%%%%%%%%%%%%%%%%%%%%%%%%%%%%%%%%%%%%%%%%%%%%%%
%  Perpendicular and parallel acceleration
%%%%%%%%%%%%%%%%%%%%%%%%%%%%%%%%%%%%%%%%%%%%%%%%%%%%%%%%%%%%
{\em Perpendicular and parallel acceleration. ---}
 Compared to balanced turbulence, for the strong imbalanced case the reduction in the energization of the particles can be seen in Fig.~\ref{fig_histograms}. Also, independently of the cross-helicity, the energy gain is significantly stronger for particle species with low values of $\ell$ (high charge-to-mass ratio) \cite{Dmitruk:2004p96}. As the electric field is all but constant on the small length scales of the gyroradius of particles with $\ell \lesssim \ell_K$, the acceleration perpendicular to the magnetic field vanishes over one gyration period and the particles are initially accelerated only by the Ohmic field in the parallel direction, $\ee_\parallel = \eta \jj_\parallel$. Due to the spontaneous formation of current sheets with large $\jj_\parallel$, the Ohmic contribution leads to a strong parallel acceleration of particles along these sheets despite the small value of the magnetic resistivity, a phenomenon previously investigated in simulations of reconnection acceleration \cite{Ambrosiano:1988p1547, Onofri:2006p39}. 

%%%%%%%%%%
\begin{figure}[b]
\begin{center}
\includegraphics[width = 0.48\textwidth]{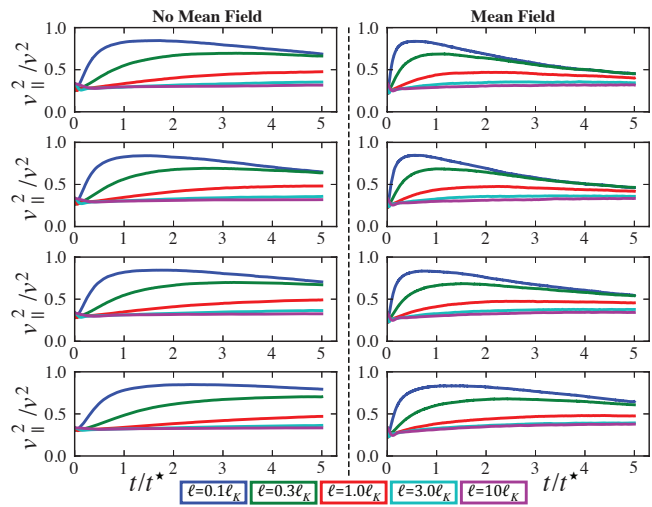}
\end{center}
\vskip -5mm
\caption{(color online). Evolution of the pitch-angle cosine square ($\alpha^2 = v_\parallel^2 / v^2$). Top to bottom $\sigma^c=\{0.0, 0.4, 0.6, 0.9\}$.}  
\label{fig_parallel}
\end{figure}
%%%%%%%%%%

However, turbulent fluctuations of the electromagnetic fields lead to pitch-angle scattering and isotropization, converting the parallel energy gained from Ohmic heating into perpendicular energy. Although slow at first, the pitch-angle scattering increases the perpendicular velocity of the particles and thus their gyroradius, which results in enhanced scattering. This effect explains the fast growth of the squared pitch-angle cosine $\alpha^2 = v_\parallel^2 / v^2$, where $v_\parallel^2$ and $v^2$ are the population-averaged squares of the velocity along the local magnetic field and the total velocity, respectively, and the subsequent decay of $\alpha^2$ back to its initial isotropic value of $0.33$ (Fig.~\ref{fig_parallel}). The pitch-angle evolution thus corresponds to the two-stage acceleration process that was observed in \cite{Ambrosiano:1988p1547}, later in \cite{Dmitruk:2004p96} and described in detail only recently in \cite{Dalena:2014p1544}.

As alignment of $\uu$ and $\bb$ reduces the intensity of the perpendicular component of the electric field, $\ee_\perp = \eta \jj_\perp - \uu \times (\BB_0 + \bb)$, the isotropization process takes longer in strongly imbalanced turbulence than in balanced turbulence. On the other hand, the presence of an external magnetic mean-field increases the pitch-angle scattering rate and results in a faster isotropization than in MHD turbulence without a magnetic mean-field.

%%%%%%%%%%
\begin{figure}[t]
\begin{center}
\includegraphics[width = 0.48\textwidth]{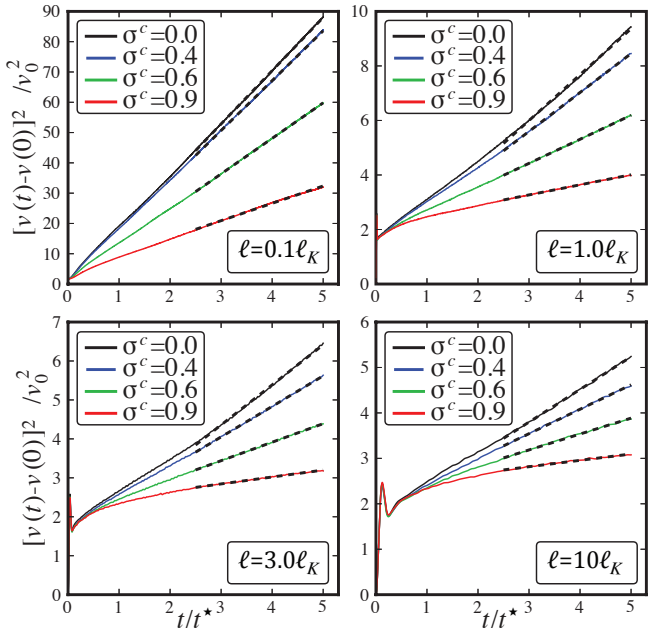}
\end{center}
\vskip -5mm
\caption{(color online). Velocity mean square displacement.}  
\label{fig_energisation}
\end{figure}
%%%%%%%%%%

%%%%%%%%%%
\begin{figure}[b]
\begin{center}
\includegraphics[width = 0.48\textwidth]{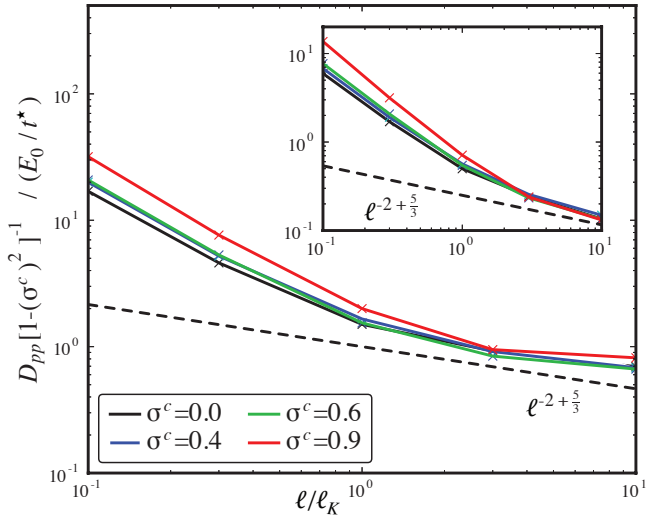}
\end{center}
\vskip -5mm
\caption{(color online). Scaling on the momentum diffusion coefficient as a function of gyroradius. The scaling takes into account the level of imbalance. Insert depicts the same figure for runs obtained in absence of a mean magnetic field.}  
\label{fig_scaling}
\end{figure}
%%%%%%%%%%

For particle species with $\ell > \ell_K$, the gyroradius is too large for the contribution of the motional electric field over one gyroperiod to be neglected. Thus $\alpha$ decreases initially, reflecting a period of perpendicular acceleration dominating over parallel acceleration, and then increases slowly as the pitch-angle distribution is isotropized again. Since the direction of the local magnetic field varies more slowly for non-zero $\BB_0$ than for $\BB_0 \equiv 0$, the initial perpendicular acceleration is much more pronounced in the runs with an external mean-field. Overall, this behavior is similar to the 'proton' case in \cite{Dmitruk:2004p96}, where a ten times stronger mean-field resulted in much faster acceleration. Consequently test-particles in our simulations stay well below the maximum energy derived in that reference.

Unlike previous studies of test-particle acceleration in time-frozen MHD turbulence \cite{Dmitruk:2003p251, Dmitruk:2004p96, Dalena:2014p1544}, our simulations implicitly include the effect of resonance between particles and propagating waves, as the MHD fields are evolved in parallel with the particle trajectories. As the MHD fields evolve in time, the effect of the dynamic fields on the trapping and entrainment of particles is modeled more realistically compared to a frozen-field approach, even though the short time acceleration, as seen in Fig.~\ref{fig_energisation}, is similar to previous works \cite{Dmitruk:2004p96} and the explicit impact of resonance is left for future works. However, the time-dependence of the fields allows us to investigate the acceleration of test-particles with a velocity similar to the Alfv\'en velocity, a regime in which resonance effects are particularly important \cite{2002cra..book.....S}, and to assess the diffusion of particles in momentum space.

%%%%%%%%%%%%%%%%%%%%%%%%%%%%%%%%%%%%%%%%%%%%%%%%%%%%%%%%%%%%
%%%%%%%%%%%%%%%%%%%%%%%%%%%%%%%%%%%%%%%%%%%%%%%%%%%%%%%%%%%%
%  Momentum diffusion estimation
%%%%%%%%%%%%%%%%%%%%%%%%%%%%%%%%%%%%%%%%%%%%%%%%%%%%%%%%%%%%
{\em Momentum diffusion estimation. ---}
 In order to estimate the effect of imbalanced turbulence on particle energization, we measure the momentum diffusion coefficient $D_{pp} = \D \langle \left[ \pp(t) - \pp(0) \right]^2 \rangle/ \D t$ (in an interval denoted by the dashed lines in Fig.~\ref{fig_energisation}). Comparing runs in simulations with varying normalized cross-helicity $\sigma^c$, we find (Fig.~\ref{fig_scaling}) that at high values of $\ell$, where we have shown $\ee_\parallel$ to be negligible, momentum diffusion scales as $D_{pp} \sim v^2_A [1-(\sigma^c)^2]$, confirming for the first time predictions from quasilinear diffusion theory \cite{Dung:1990p1533}, which argue that, for a scattering time which depends on the spectral properties of the turbulence,
\begin{align}
D_{pp} \sim \frac{p^2}{4 \tau} \frac{v_A^2}{v^2} \frac{\delta b^2}{B_0^2} \left[ 1-\left(\sigma^c\right)^2 \right].
\end{align}
The scattering time $\tau$ depends on the gyroradius, gyro-frequency, and the spectral properties of the turbulence. For isospectral slab turbulence (for which all the ideal invariants are taken to have the same power law scale dependence), with spectral index $s$ and a mono-energetic particle distribution, the model of Dung and Schlickeiser predicts $\tau \sim \ell^{2-s}$. This agrees surprisingly well with our DNS simulations at larger values of $\ell$ if we take $s=5/3$, especially considering the anisotropy of our directly simulated MHD turbulence is at odds with Dung and Schlickeiser's turbulence model.

%%%%%%%%%%%%%%%%%%%%%%%%%%%%%%%%%%%%%%%%%%%%%%%%%%%%%%%%%%%%
%%%%%%%%%%%%%%%%%%%%%%%%%%%%%%%%%%%%%%%%%%%%%%%%%%%%%%%%%%%%
%  Discussion and conclusions
%%%%%%%%%%%%%%%%%%%%%%%%%%%%%%%%%%%%%%%%%%%%%%%%%%%%%%%%%%%%
{\em Discussion and conclusions. ---}  Using numerical simulations of MHD turbulence, we observed that the energization of Alfv\'enic test particles ($v\approx v_A$) for balanced turbulence is more pronounced compared to a strongly imbalanced state. This is indicative of a systematic acceleration loss affecting the particles in the imbalanced case. This implies a weaker ion heating rate in plasmas characterized by strong imbalance, such as the fast solar wind. For fast particles ($v\gg v_A$), for which the electric field acceleration is small, the spatial diffusion is not expected to vary with the imbalance, as reported by Ref. \cite{Beresnyak:2011p1446}. However, on the basis of our numerical simulations we demonstrate that particle acceleration and thus momentum diffusion is very sensitive to cross helicity. Our results generalize earlier findings based on the quasilinear approximation \cite{Dung:1990p1557,Dung:1990p1533} to particle dynamics.

In our study, compared to the smallest scale of turbulence ($\ell_K$), particles with various initial gyro-radii are selected. As the particles are non-adiabatic, the gyro-radius will change in time and the particles will resonate with different Alfv{\'e}n wave wavelengths, of different energies. Gyro-radii of the order of the Kolmogorov scale and smaller give particles a stronger adiabatic characteristic. While smaller gyro-radii particles experience more efficient heating, the tendency of imbalanced turbulence to suppress the acceleration is shown to be present at all scales. This result implies a need for kinetic simulations to account for the level of imbalance of the larger plasma scales that act as an energy source in the system.

In Ref.~\cite{Chandran:2010p1532} it was conjectured that the perpendicular heating rate of ions due to low-frequency Alfv\'en waves with wavelengths on the scale of the ion gyroradius was independent of the degree of imbalance of the turbulence, as long as the root-mean-square fluctuations of the velocity and the magnetic field stay constant. The results of this letter imply that, on the contrary, the heating rate is strongly reduced in imbalanced turbulence.

The research leading to these results has received funding from the European Research Council under the European Union's Seventh Framework Programme (FP7/2007--013)/ERC Grant Agreement No. 277870 and Deutsche Forschungsgemeinschaft priority program 1573 (Physics of the Interstellar Medium).

%%%%%%%%%%%%%%%%%%%%%%%%%%%%%%%%%%%%%%%%%%%%%%%%%%%%%%%%%%%%%%
%%%%%%%%%%%%%%%%%%%%%%%%%%%%%%%%%%%%%%%%%%%%%%%%%%%%%%%%%%%%%%
%\bibliography{Reference_Alignment_Accelerate_PoP}


%%%%%%%%%%%%%%%%%%%%%%%%%%%%%%%%%%%%%%%%%%%%%%%%%%%%%%%%%%%%%%
%%%%%%%%%%%%%%%%%%%%%%%%%%%%%%%%%%%%%%%%%%%%%%%%%%%%%%%%%%%%%%
\end{document}